\documentclass[fleqn,10pt]{wlscirep}
\usepackage{amsfonts}
\usepackage{graphicx}
\usepackage{mathtools}
\usepackage{longtable}
\usepackage{float}
\usepackage{soul}
\usepackage{xcolor}

\title{The nested structural organization of the worldwide trade multi-layer network}

\author[1,*]{Luiz G. A. Alves}
\author[2]{Giuseppe Mangioni}
\author[3]{Isabella Cingolani}
\author[1,8,9]{Francisco Aparecido Rodrigues} 
\author[4]{Pietro Panzarasa}
\author[5,6,7]{Yamir Moreno}

\affil[1]{Institute of Mathematics and Computer Science, University of S\~ao Paulo, S\~ao Carlos, SP 13566-590, Brazil}
\affil[2]{Dipartimento di Ingegneria Elettrica, Elettronica e Informatica, University of Catania, Catania 95125, Italy}
\affil[3]{Big Data and Analytical Unit, Department of Surgery and Cancer, Imperial College London, London SW7 2AZ, UK}
\affil[4]{School of Business and Management, Queen Mary University of London, London E1 4NS, UK}
\affil[5]{Department of Theoretical Physics, University of Zaragoza, Zaragoza 50009, Spain}
\affil[6]{Institute for Biocomputation and Physics of Complex Systems, University of Zaragoza, Zaragoza 50009, Spain}
\affil[7]{ISI Foundation, Torino 10126, Italy}
\affil[8]{Mathematics Institute, University of Warwick, Gibbet Hill Road, Coventry CV4 7AL, UK}
\affil[9]{Centre for Complexity Science, University of Warwick, Coventry CV4 7AL, UK}

\affil[*]{lgaalves@usp.br}
\affil[+]{All authors contributed equally to this work}


\begin{abstract}
Nestedness has traditionally been used to detect assembly patterns in meta-communities and networks of interacting species. Attempts have also been made to uncover nested structures in international trade, typically represented as bipartite networks in which connections can be established between countries (exporters or importers) and industries. A bipartite representation of trade, however, inevitably neglects transactions between industries. To fully capture the organization of the global value chain, we draw on the World Input-Output Database and construct a multi-layer network in which the nodes are the countries, the layers are the industries, and links can be established from sellers to buyers within and across industries. We define the buyers' and sellers' participation matrices in which the rows are the countries and the columns are all possible pairs of industries, and then compute nestedness based on buyers' and sellers' involvement in transactions between and within industries. Drawing on appropriate null models that preserve the countries' or layers' degree distributions in the original multi-layer network, we uncover variations of country- and transaction-based nestedness over time, and identify the countries and industries that most contributed to nestedness. We discuss the implications of our findings for the study of the international production network and other real-world systems.

\end{abstract}

\begin{document}

\flushbottom
\maketitle
\thispagestyle{empty}

\section*{Introduction}

Nestedness was originally proposed to uncover biogeographic meta-community patterns of occurrence of species and patterns of interaction among species in mutualistic ecological networks in which mutually beneficial interactions occur between participants of two distinct sets~\cite{atmar1993,almeida2008consistent,bascompte2003nested,bastolla2009architecture,patterson1986}. Typically, nestedness has been used to capture the extent to which more specialist species interact with proper subsets of species that, in turn, interact with more generalist ones. In ecological mutualistic networks, the nested architecture has been shown to minimize competition between species, and therefore to enable the system to foster greater biodiversity~\cite{bastolla2009architecture}. In particular, because in a nested system specialist species tend to interact only with generalist species, and because the latter tend to be less vulnerable than the former, nestedness is expected to amplify the chances of survival of rare species. A recent work, however, has suggested that decreased risk of extinction is distributed heterogeneously across the nodes of ecological networks. In particular, evidence has shown that there is a negative relationship between nodes' contributions to the nested architecture of the system and their individual survival benefits~\cite{saavedra2011strong}. 

Applications of nestedness beyond biological ecosystems are not new~\cite{Erman,konig,leontief1965structure,saavedra2009simple,saavedra2011strong}. For instance, recent studies have suggested that the trajectories followed by the productive structures of countries and regions tend to be shaped by the underlying product space in which any two products are connected if they are exported by two or more countries~\cite{hidalgo2007}. Similarly, it has been suggested that the structure of the taxonomy network between products resulting from the export of countries is associated with countries' potential growth and development paths~\cite{zaccaria2014taxonomy}. Moreover, recent work has concentrated on the nestedness of economic systems to shed light on the economic geography of domestic and international trade~\cite{bustos2012dynamics}. In particular, it has been suggested that industry--location networks display a nested structure that tends to remain stable over time and can help predict the evolution of countries' product space and industrial reconfigurations. Since the diversity of products that countries export has direct bearing on economic growth~\cite{bustos2012dynamics}, understanding the underlying nested structure emerging from trade can therefore help design more effective policies aimed at strengthening and sustaining countries' economic prosperity. 

Traditionally, scholars have investigated a system's nestedness by formalizing the bipartite networks in which a node belonging to a group is assumed to be linked with another node in a different group if there is an interaction between them~\cite{dormann2009indices,thebault2010stability,fortuna2010nestedness,jonhson2013factors,staniczenko2013ghost,fortuna2010nestedness,cristelli2013measuring,saracco2015randomizing}. This has also been the case with most empirical studies of the international trade between countries~\cite{bustos2012dynamics,Erman,konig}. Typically, in a bipartite trade network, a set of nodes represents the countries and another set includes the industries to (from) which the countries export (import)~\cite{bustos2012dynamics}. However, a bipartite network connecting countries to (exported or imported) products cannot account for the full extent of interactions that typically occur among countries in the international production network underpinning the global value chain.

Indeed, countries are traditionally involved in economic transactions within and across multiple industries. In addition, a large number of transactions can originate from, and terminate at, the same country, both within and across industries, thus contributing to the various production stages along which intermediary products are transformed into final ones~\cite{cingolani}. A bipartite network connecting countries to products would be unable to fully represent all such transactions. It neglects the possible transactions within and across industries, and does not disentangle transactions within the same country from those occurring between different countries. Even if we considered the one-mode projections of the bipartite network~\cite{saracco2015randomizing,cristelli2013measuring,mastrandrea2014reconstructing}, such as the product-to-product or country-to-country networks, we would only be able to focus on one type of interaction at a time, and in any case we would be unable to assess the assembly patterns among the various productions stages to which countries contribute. Moreover, even when trade networks are formalized as multiplex networks~\cite{menichetti}, in which the nodes are the countries and the layers are the industries in which countries trade with one another, connections between countries would only be allowed within layers, and any involvement of countries in relationships between industries would therefore be neglected.

Transactions of goods or services between industries within and across countries are related to the production stages into which the global value chain can be articulated~\cite{cingolani}. Thus, to properly evaluate the nestedness of countries with respect to production stages as well as the nestedness of production stages with respect to countries, a more comprehensive approach to trade would be needed that combines both international and domestic transactions occurring within and across the various stages of the global value chain. In this work, we take a step in this direction and investigate economic nestedness using a multi-layer representation of the worldwide production network~\cite{boccaletti,kivela}. 

To build this multi-layer network, we consider the World Input-Output Database (WIOD)~\cite{timmer2015illustrated}, covering data on exchange of intermediate and final products and services among $43$ countries and $57$ different economic activities (industries) in the period from 2000 to 2014. In this multi-layer network, each economic activity is represented as a layer, and each layer is populated by the $43$ countries in our data set. Connections are directed from sellers (i.e., countries selling a product or a service) to buyers (i.e., countries purchasing a product or service). A connection is established between any two countries when there is an economic transaction between them either within the same industry (intra-layer connections) or across different economic industries (cross-layer connections)~\cite{timmer2015illustrated}. Moreover, in this multi-layer network, a given country can exchange products or services to itself when a transaction takes place from one industry to another or within the same industry, and the same country is involved both as a seller and as a buyer. 

Based on the multi-layer network, to assess nestedness we construct the buyers' and sellers' participation matrices in which buyers' and sellers' involvement in transactions within and across layers can be measured. Our findings suggest that the nested structure of these matrices is similar to the one uncovered in ecological networks~\cite{atmar1993,patterson1986}. Unlike other studies of nestedness based on bipartite trade networks~\cite{bustos2012dynamics,Erman} or one-mode projections of bipartite networks~\cite{konig}, we draw on countries' involvement in the various stages of the global value chain, and investigate the nestedness of countries (with respect to production stages) and of production stages (with respect to countries). We show how values of country- and transaction-based nestedness vary over time, and distinguish between the cases in which suppliers or buyers are involved in the transactions. To assess the statistical significance of our findings, we compare the actual values of country- and transaction-based nestedness with the ones obtained using appropriate null multi-layer models in which links are reshuffled while the countries' or layers' degree distributions, respectively, are preserved. We further evaluate how individual countries and individual industries contribute to nestedness by drawing on null models in which the connections of each country or each industry, respectively, are reshuffled one at a time. We then argue in favor of the salience of our results for the study of economic stability and growth as well the system's vulnerability to exogenous shocks. Finally, because multi-layer networks can be found across a variety of biological, technological and social systems, we discuss the implications that our proposed approach to measuring nestedness can have beyond trade, for a wide range of empirical domains. 

\section*{Results}

\subsection*{The data set}

Our study draws on data from the WIOD (Release 2016) covering $28$ EU countries and $15$ other major countries in the world within the period from 2000 to 2014. For every year, a World Input-Output Table (WIOT) is provided in current prices, expressed in millions of US dollars (USD). Each table represents economic exchanges among the $56$ economic activities (industries) in each country and their respective final demand. The final demand is represented by five separated components: the final consumption expenditure, the final consumption expenditure by non-profit organizations serving households (NPISH), the final consumption expenditure by government, the gross fixed capital formation, and the changes in inventories and valuable. For the purpose of this work, the five final demand components were combined into a unique aggregated component representative of all product consumption (i.e., individuals, non-profit organizations, and enterprises), capital formation, governmental expenditure, and changes in inventories. 

\begin{figure}[!ht]
\centering
\includegraphics[width=0.95\textwidth]{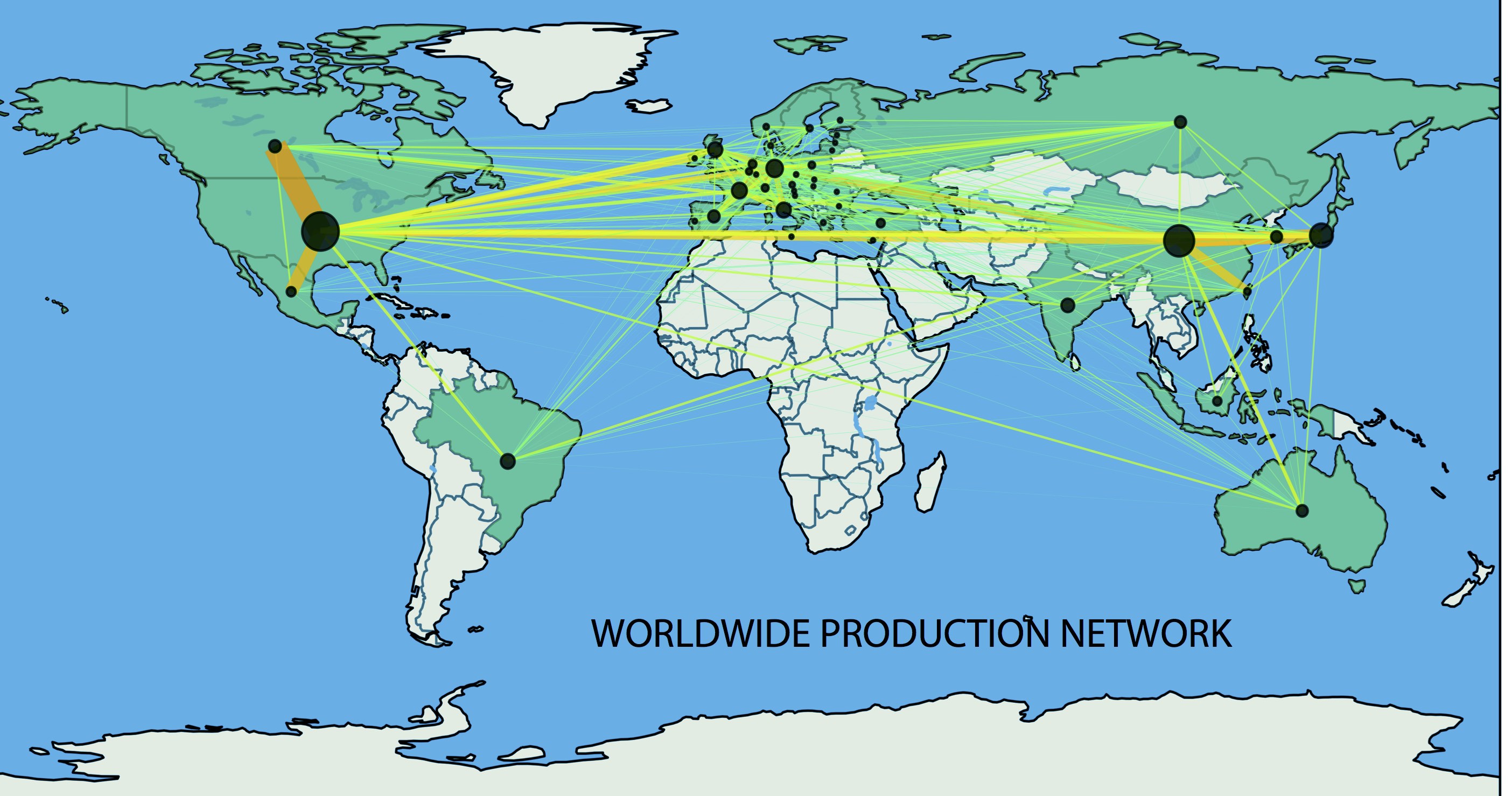}
\caption{Network of the overall worldwide trade network in 2010. Network representation of the total amount exchanged between and within countries across all $57$ economic industries in 2010. The shade of the links between countries as well as their widths are proportional to the total amount (in millions of US dollars) exchanged between the connected countries. The darker the color, the greater the amount exchanged between the countries. Each node's size is proportional to the value exchanged and/or consumed within the corresponding country.}
\label{fig:map}
\end{figure}

Fig.~\ref{fig:map} shows a network representation of the aggregated economic interactions among countries in all sectors of activity in 2010. The nodes of the network in Fig.~\ref{fig:map} represent the countries, links refer to trade, and the intensity of the color of links between countries as well as their width vary as a function of the total amount of value exchanged between the connected countries across all economic activities. Finally, each node's size in Fig.~\ref{fig:map} reflects the value exchanged within the corresponding country. Indeed the international production network comprises both a domestic and a strictly international trade component. Thus, to account for this, the size of each node was made proportional to the sum of the corresponding country's internally exchanged products/services, i.e., the sum of the value of all intermediate exchanged inputs and/or consumption within the country.

Notice that, while the WIOD provides the most complete publicly available representation of trade between countries and industries on a large international scale, the data set is restricted to a limited sample of countries and industries. On the other hand, while other available data sets account for more countries and industries (e.g., the United Nations COMTRADE data), they do not provide as detailed information as the WIOD on single economic transactions, and cannot therefore be used for the analysis of the global value chain. Thus, the price of using a more detailed description of trade for studying the international production network is paid in terms of a coarse-grained description of the industrial sectors and of the incompleteness of economic transactions, which in principle may affect the results.

To address the shortcomings of using bipartite networks to study nestedness, in what follows we shall propose an approach based on a multi-layer trade network in which every layer represents an economic industry (i.e., a set of products or services classified as similar given their nature and economic function), populated by the trading countries that, unlike what happens in a multiplex network, can now trade with themselves or with other countries both within and across layers. Notice that, in what follows, we shall use the terms ``product'' (or ``service''), ``economic activity'', and ``industry'' interchangeably to refer to a single layer of the network, which in turn may represent either a single (intermediary or final) production stage at which an economic transaction can occur or the final consumption.

\subsection*{The multi-layer trade network}
Based on data from the WIOD, we start by constructing the multi-layer trade network reproducing the complex system of transactions within and between industries and within and between countries. To this end, we obtain a block matrix  including: (i) $57$ diagonal sub-matrices, each referring to transactions within one single layer (i.e., 56 sub-matrices representing the economic activities showed in Table~S1 and one sub-matrix representing the sum of the five components resulting in the final demand); (ii) $3,192$ off-diagonal matrices representing transaction betweens pairs of distinct layers. Each of the square diagonal and off-diagonal sub-matrices has 43 rows (columns) corresponding to the countries showed in Table~S2. 

\begin{figure}[!ht]
\centering
\includegraphics[width=0.95\textwidth]{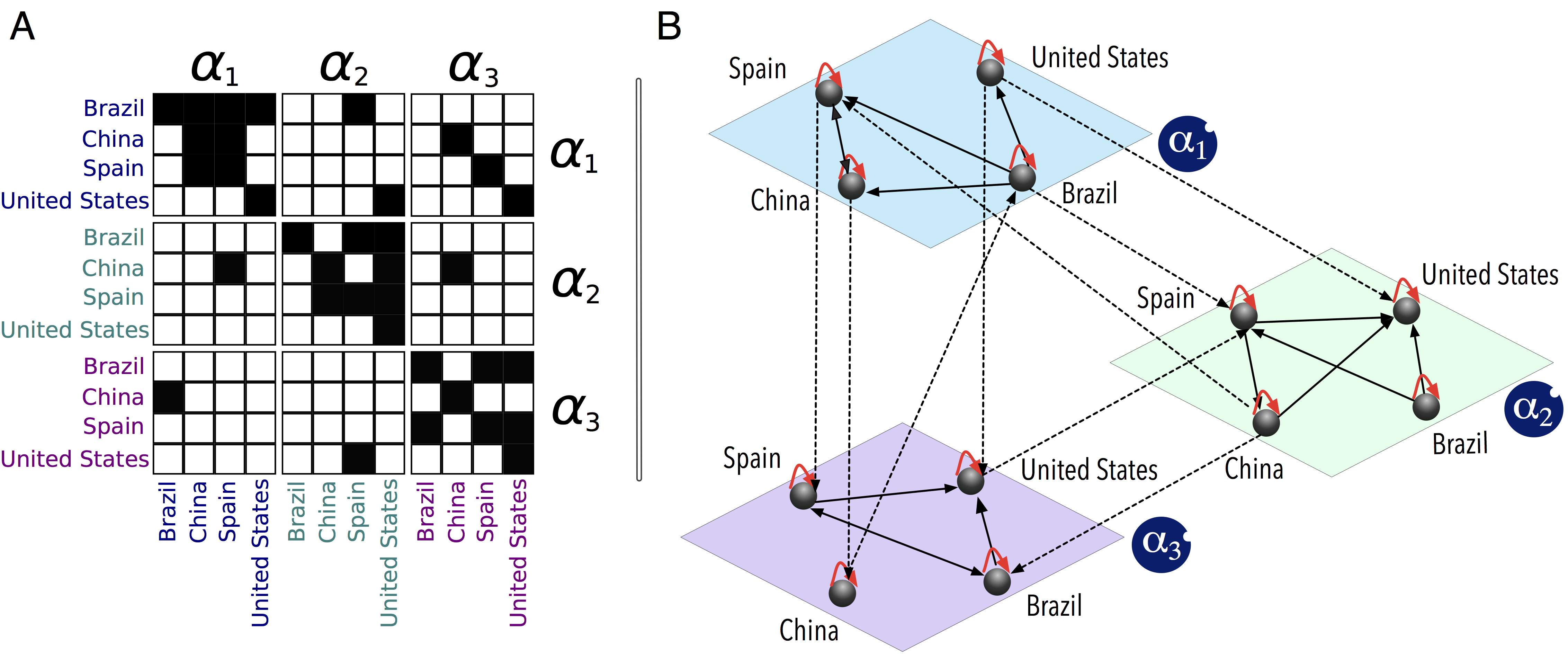}
\caption{Schematic representation of a multi-layer trade network. A) The adjacency matrix containing four countries $c_i$ and three hypothetical products (layers) $\alpha_i$. The three diagonal sub-matrices represent the connections among countries in the same layer, whereas the other off-diagonal sub-matrices represent the cross-layer connections among countries. B) Visual representation of the corresponding multi-layer network built from the matrix. The cross-layer connections are represented by dashed lines and the intra-layer connections by solid ones. For both types of connections, the arrow represents the directionality of each economic exchange originating from a seller and pointing to a buyer. Self-loops of length one represent transactions occurring within the same country and the same layer.}
\label{fig:multilayer}
\end{figure}

Thus, each cell in the resulting block matrix provides the USD values of products and services exchanged within and across aggregated economic activities (industries) and within and across countries. The block matrix is very dense and visually little informative. For this reason and for illustrative purposes, Fig.~\ref{fig:multilayer}A shows a simplified matrix displaying the data provided by the WIOD on a reduced scale. The matrix in Fig.~\ref{fig:multilayer}A shows the transactions among four countries $c_i$ concerned with three hypothetical industries $\alpha_i$. A cell $a^{\alpha_i}_{c_i,c_j}$ is black if there is a transaction (i.e., the USD value exchanged is different from zero) between country $i$ and country $j$ within industry $\alpha_i$, and white if there is no such transaction (i.e., value exchanged equal to zero). Notice that transactions are not symmetric, and thus the buyers (columns) and sellers (rows) are likely to play different roles in the structural organization of the worldwide production network. Fig.~\ref{fig:multilayer}B provides a visual representation of the adjacency matrix reported in Fig.~\ref{fig:multilayer}A. Notice that the three-layer network includes: (i) transactions between different countries within the same industry, represented by the intra-layer connections (e.g., transactions from Brazil to China in layer $\alpha_1$); (ii) transactions between different countries across different industries, represented by the cross-layer connections (e.g., transactions from Brazil in layer $\alpha_1$ to Spain in layer $\alpha_2$); (iii) transactions across industries involving the same country, represented by cross-layer connections departing from and point to the same node (e.g., transactions from layer $\alpha_1$ to layer $\alpha_2$ from and to the United States); and (iv) transactions within the same industry and involving the same country, represented by self-loops of length one (e.g., transactions in layer $\alpha_1$ from and to the United States). Notice that these self-loops are displayed as black diagonal cells in the diagonal matrices in Fig.~\ref{fig:multilayer}A, and as arrows departing from and pointing to the same node in Fig.~\ref{fig:multilayer}B.

We define a multi-layer network as a pair $M=(G,C)$, where $G=\{G_{\alpha};~\alpha~\in~\{1,\dots,k\}\}$ is a family of directed graphs $G_{\alpha}=(X_{\alpha},E_{\alpha})$ called layers of $M$, and $C$ is the set of interconnections between nodes belonging to different layers $G_{\alpha}$ and $G_{\beta}$ with $\alpha \neq \beta$. Formally,
\begin{equation}
C = \{ E_{\alpha \beta} \subseteq X_\alpha \times X_\beta;\; \alpha,\beta \in \{ 1,\dots,k\}, \alpha \neq \beta \}.
\end{equation}
The elements of $C$ are called ``cross--layer connections'', and the elements of each $E_\alpha$ are called ``intra--layer connections''. On the one hand, given a layer $G_\alpha$ corresponding to one of the $56$ economic industries or the final demand, the $N_\alpha=43$ nodes corresponding to the countries are denoted by $X_\alpha=\{ c_1^\alpha,\dots,c_{N_\alpha}^\alpha  \}$, and the intra-layer adjacency matrix of each layer $G_\alpha$ will be denoted by $A^{[\alpha]}=(a_{ij}^\alpha)$, where:
\begin{equation}
\begin{array}{c}
a_{ij}^\alpha =
  \begin{cases}
    1  & \quad \text{if } (c_i^\alpha, c_j^\alpha) \in E_\alpha\\
    0  & \quad \text{otherwise},\\
  \end{cases}\\
\end{array}
\end{equation}
\noindent for $1 \leq i, \; j \leq N_\alpha$  and $1 \leq \alpha \leq k$. An intra-layer connection is established from node $i$ to node $j$ in layer $\alpha$ if there is at least one economic exchange from country $c_i$ to country $c_j$ in the same layer $\alpha$. Notice that $a_{ii}^\alpha=1$ would refer to transactions occurring within the same country $c_i$ and the same layer $\alpha$. On the other hand, the cross--layer adjacency matrix corresponding to $E_{\alpha \beta}$ is the matrix $A^{[\alpha,\beta]}=(a_{ij}^{\alpha \beta})$ given by:
\begin{equation}
a_{ij}^{\alpha \beta} =
  \begin{cases}
    1  & \quad \text{if } (c_i^\alpha, c_j^\beta) \in E_{\alpha \beta}\\
    0  & \quad \text{otherwise.}\\
  \end{cases}
\end{equation}
A cross-layer connection is established from node $i$ in layer $\alpha$ to node $j$ in layer $\beta$ when there is at least one economic exchange from country $c_i$ in layer $\alpha$ to country $c_j$ in layer $\beta$. Once again, $a_{ii}^{\alpha \beta}=1$ would imply that a transaction occurs within the same country $c_i$ from layer $\alpha$ to layer $\beta$.

\subsection*{Nestedness of countries and products}

\begin{figure}[!h]
\centering
\includegraphics[width=0.95\textwidth]{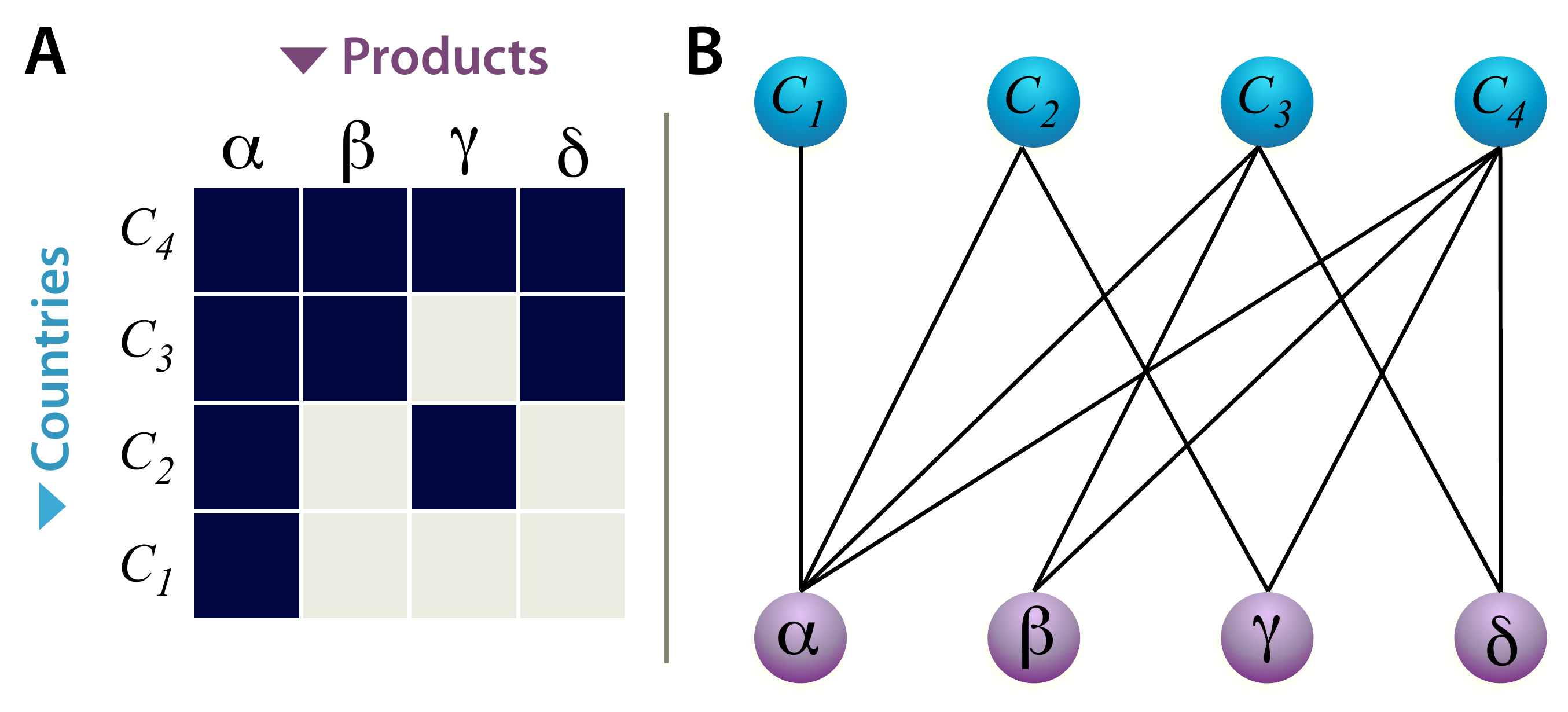}
\caption{Nestedness of country-product connections. A) Matrix representation of the connections between countries and exported products ordered by degree centrality. B) Bipartite network of country-product connections. Countries are hierarchically organized starting from the one ($c_1$) that exports products that all other countries export. Notice that such bipartite network representation does not distinguish between transactions occurring within the same country and transactions involving different countries. This network representation is also unable to distinguish between transactions across different industries and transactions within the same industry, and therefore does not capture the full organization of the global value chain.}
\label{fig:bipartite}
\end{figure}

In economic systems, nestedness is akin to maximal possible diversification subject to the constraints of international competition. For instance, in the simplified case of international trade between countries with no domestic intra- and inter-layer exchange, an economic system can be regarded as nested when a number of countries export (import) a proper subset of the products exported (imported) by other countries, which in turn export (import) a (larger) proper subset of the products exported (imported) by other countries, and so forth (see Fig.~\ref{fig:bipartite}). Countries can, therefore, be hierarchically organized into progressively richer levels such that as countries move from an inner to an outer level they are involved in the trade of more products. Notice that, unlike the simplified case of international trade in this example, our multi-layer perspective enables us to capture the intricacies of both domestic and international trade as well as countries' involvement in multiple stages of the global value chain.  
 
Moreover, a nested structure of the bipartite country-product network would imply that countries in outer levels (e.g., country $c_4$ in Fig.~\ref{fig:bipartite}) are less similar to countries in inner levels (e.g., country $c_1$ in Fig.~\ref{fig:bipartite}) with respect to the products traded than vice versa~\cite{tversky}. As a result, the countries belonging to the core of the system are those associated with the largest degree of similarity to all other countries with respect to the products exported (imported). Yet, while the countries in the core of the system are connected to fewer products than the countries in outer levels, the former do not necessarily concentrate their exports (imports) on any of the products they trade. Indeed their export (import) profile may be characterized by a homogeneous distribution of trade across a (relatively small) number of products. On the other hand, the countries in the outermost level are those with the most diverse trade profile in the system. That is, they are connected to all products traded in the system, and among these products there is at least one of which they are the sole traders. However, while these countries in the outermost level are connected to more products than the countries in inner levels, they do not necessarily spread their efforts uniformly across products. Indeed, among the many products they trade, they may well concentrate most of their economic transactions on a select minority of them.

The idea of hierarchically organizing the elements of an economic system into progressively richer subsets can also be applied to (imported or exported) products. In particular, the core products of a system are the smallest number of products imported (exported) by the largest number of countries. In turn, the core products represent a proper subset of the products imported (exported) by other countries, and so forth up to the final set of products imported (exported) by the countries in the outermost level. As products move from an inner to an outer level, they are traded by fewer countries and only in larger combinations with other products. At one extreme, an economy may be organized in such a way that all products except one are traded by all countries, while the remaining product is traded only by one country. In this case, the only product controlled by one country occupies the outer level, while all others belong to the core (most nested part) of the system.

Core products, being the ones with the highest country-level substitutability, are likely to be based on the most widespread know-how, technologies, and competences. Less nested and more ``peripheral'' products, on the other hand, are likely to be more country-specific and characterized by lower degrees of country-level substitutability. It may be speculated that, if low degrees of product nestedness may secure high returns to the producing countries, it may also prompt a large degree of market instability should an external shock affect the few countries that are the sole suppliers of the product. 

\subsection*{Nestedness in the multi-layer network}

We now extend the notion of nestedness so far discussed in connection with a bipartite network to account for the complexity of a multi-layer network in which transactions can be both domestic and international and can originate from, and point to, different industries within different countries. More specifically, unlike the simplified case of the bipartite country-product network, in a multi-layer trade network a given country is not simply connected to products but, more properly, it buys or sells products within specific transactions that take place from an industry to another or even within the same industry. Moreover, these transactions may have different countries as suppliers and buyers, or they may even occur within the same countries. The multi-layer perspective, therefore, enables us to shift focus from products (industries) to transactions, and to draw on these transactions to evaluate the nestedness of both countries and production stages within the global value chain. For instance, the core suppliers can be defined as the largest set of countries involved as sellers in the smallest number of production stages in which all other countries are involved as suppliers. By contrast, the supplier in the outermost level would be the one involved as a seller in at least one production stage in which no other country is involved. Similarly, we can assess the nestedness of production stages in the global value chain, and measure the degree to which country-poor production stages are proper subsets of country-rich ones. 

\begin{figure}[!ht]
\centering
\includegraphics[width=0.92\textwidth]{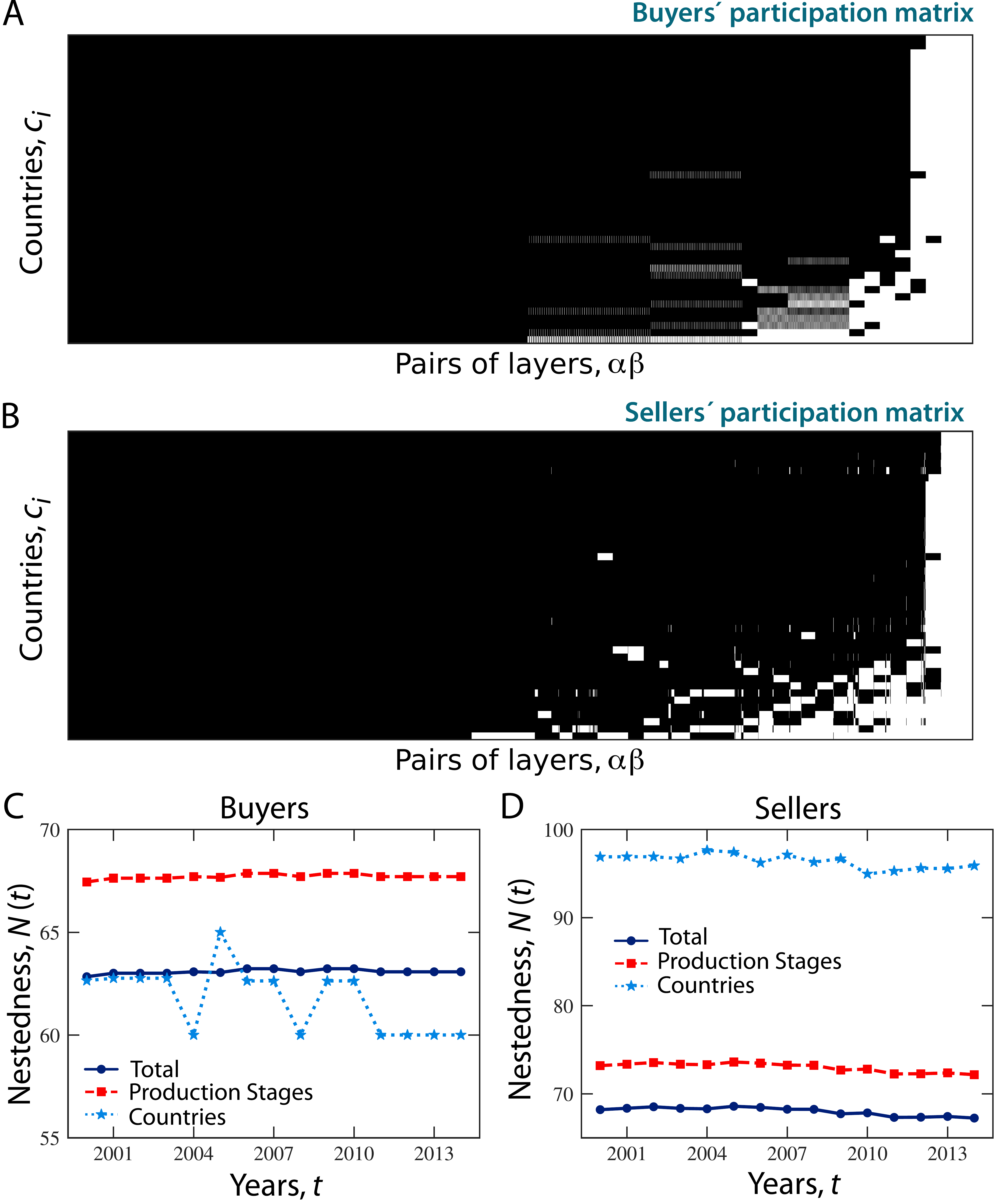}
\caption{Nested organization of buyers' and sellers' participation matrices. Buyers' (A) and sellers' (B) participation matrices in 2010. A generic element $B_{c,\alpha\beta}$ ($S_{c,\alpha\beta}$) of the matrix is equal to $1$ (black square) if there is a link ending (starting) at (from) country $c_i$ in layer $\alpha$ and starting (ending) from (at) layer $\beta$, and is equal to zero otherwise (white square). In a perfectly nested matrix, the black cells should fill in the upper triangular portion of the matrix (above the secondary diagonal) and the white cells should lie in the lower one (below the secondary diagonal). (C-D) Nestedness calculated over the years from 2000 to 2014 for buyers' (C) and sellers' (D) participation matrices.} 
\label{fig:nestedness}
\end{figure}

Before we can compute nestedness using the WIOD, we need to build a matrix, called the \textit{participation matrix}, in which rows are the countries and columns are all possible \textit{ordered combinations of any two layers} (i.e., productions stages or economic transactions between or within industries). This means that, for every year, we have a participation matrix with $43$ rows and $57 \times 57 = 3,249$ columns. Moreover, in each layer, we can distinguish between buying countries, i.e., countries with incoming links, and selling countries, i.e., countries with outgoing links. Thus, for every year we built two matrices -- the buyers' and the sellers' participation matrices -- where a generic element, $B_{c_i,\alpha\beta}$ or $S_{c_i,\alpha\beta}$, is equal to $1$ if there is a link, respectively, ending at or starting from country $c_i$ in layer $\alpha$ and starting from or ending at layer $\beta$, and zero otherwise. Figs.~\ref{fig:nestedness}A and B show the buyers' and sellers' participation matrices, respectively, in which the black color refers to a value of $1$ in the corresponding matrix, while the white color refers to zero.

Drawing on the buyers' and sellers' participation matrices, we computed the nestedness of the multi-layer network. To this end, first we reordered the columns and rows by decreasing degree centrality, i.e., respectively, the number of countries participating in transactions, and the number of transactions in which countries are involved~\cite{jonhson2013factors,dominguez2015ranking}. We then computed nestedness by using the measure proposed by Almeida-Neto \textit{et al.}~\cite{almeida2008consistent}, referred to as \textit{NODF} and here denoted by $N$. More specifically, this measure $N$ is based on two properties: decreasing fill ($DF$) and paired overlap ($PO$). Let us suppose that the above defined participation matrices have $m$ rows and $n$ columns and consider a pair of rows $(i,j)$ such that $i < j$ and a pair of columns $(k,l)$ with $k < l$. Let $MT$ be the marginal total (i.e., the sum of ones) of any column or row. For any pair $(i,j)$ of rows, $DF_{ij}$ is defined as equal to $100$ if $MT_j < MT_i$ and zero otherwise. Similarly, for any pair of columns $(k,l)$, $DF_{kl}$ is equal to $100$ if $MT_l < MT_k$ and zero otherwise.\\

The paired overlap can be computed as follows. For any pair of columns $(k,l)$ such that $k < l$, $PO_{kl}$ is the percentage of ones in column $l$ that are located at the same row positions as the ones in column $k$. Similarly, for any two rows $(i,j)$ such that $i < j$, $PO_{ij}$ is the percentage of ones in row $j$ that are located at the same column positions as the ones in row $i$. Formally, given any left-to-right column pair and any up-to-down row pair, the degree of paired nestedness ($N_{paired}$) is defined as
\begin{equation}
N_{paired} = 
\begin{cases} 
   0 & \text{if } DF_{paired}=0; \\
   PO  & \text{if } DF_{paired}=100.
  \end{cases}
\end{equation}
The measure of row (column) nestedness $N_{row}$ ($N_{col}$) is calculated by averaging all values of paired row (column) nestedness. Notice that the total number of values of row and column paired nestedness for a matrix with $n$ rows and $m$ columns is $n(n-1)/2$ and $m(m-1)/2$, respectively. Thus, we define the nestedness of the whole matrix as
\begin{equation}
\centering
N = \frac{\sum N_{paired}}{\frac{n(n-1)}{2} + \frac{m(m-1)}{2}}.
\end{equation}
So conceived of, the values of nestedness range between $0$ and $100$.

In our case, the total nestedness among rows is a measure of nestedness of countries with respect to production stages (i.e., economic transactions here defined as combinations of products or industries). It refers to the degree to which a subset of countries are involved in transactions between industries that are a proper subset of the more diverse transactions in which other countries are involved, and so forth. That is, the countries controlling country-poor stages of production (or transactions) constitute proper subsets of the countries involved in country-rich stages of production. Thus, in a perfectly (country-based) nested economic system, the core country would be the one involved in the smallest set of transactions within the international production network in which all other countries are involved. The most peripheral country lying in the outermost level, by contrast, would be the one involved in all transactions in which all other countries are involved and in at least one transaction in which no other country is involved. Notice that to have a perfectly (country-based) nested economic system, no pair of countries can be involved in the same (number of) production stages; yet, any pair of countries may differ by more than one associated production stage.

On the other hand, nestedness among columns quantifies the nestedness of stages of production or economic transactions with respect to countries. This refers to the degree to which a number of stages of production involve countries that constitute a proper subset of the countries involved in other production stages, and so forth up to the stage at which all countries are involved. Similarly, in a perfectly (transaction-based) nested system, the core production stage would be the one in which all countries are involved, while the most peripheral transaction lying in the outermost level would be the one controlled only by one country. Once again, to have a perfectly (transaction-based) nested system, no pair of production stages or transactions can be controlled by the same (number of) countries; yet, any two transactions may differ by more than one involved country. 

Finally, the matrix nestedness is a measure of the nestedness of the whole multi-layer trade network in a given year. It thus combines country-based and transaction-based nestedness. In this sense, a perfectly nested economic system would be both perfectly country-based and transaction-based nested. Thus, no pair of countries can be involved in the same (number of) production stages, and no pair of production stages can be controlled by the same (number of) countries. In addition, any two adjacent rows (countries) may differ only by one production stage, and any two adjacent columns (productions stages) may differ by only one country. 

In our multi-layer network, each country can be the buyer or the supplier in each production stage, and therefore the measures of nestedness outlined above can be computed both for buying and for selling countries, in each year. Buyers' country-based nestedness refers to the degree to which the countries that act as buyers (of intermediary or final products) in buyer-poor stages of production constitute proper subsets of the countries that act as buyers in buyer-rich stages of production. Similarly, sellers' country-based nestedness refers to the degree to which the countries that act as suppliers in seller-poor stages of production constitute proper subsets of the countries acting as suppliers in seller-rich production stages. Finally, transaction-based nestedness from the buyers' (sellers') perspective refers to the degree to which stages of production in which few buyers (suppliers) are involved constitute proper subsets of the production stages in which more buyers (suppliers) are involved. 

Fig.~\ref{fig:nestedness} shows the evolution over time of both country-based and transaction-based nestedness from the perspective of both buyers (Fig.~\ref{fig:nestedness}C) and sellers (Fig.~\ref{fig:nestedness}D). Results suggest that, while sellers' country-based nestedness remained constant over the years, buyers' country-based nestedness fluctuated between 2003 and 2011. Moreover, sellers' nestedness remained higher than buyers' nestedness constantly over the years. In particular, sellers' country-based nestedness was remarkably higher than buyers' country-based nestedness, thus suggesting a more structured organization of countries' involvement in production stages in which countries acted as sellers of intermediate and final products than in stages where countries acted as buyers. Finally, it is worth noting that the ordering between country-based nestedness and transaction-based nestedness varies depending on the role of countries as buyers or sellers. When countries participated in transactions as buyers, the nestedness of production stages with respect to countries was larger than the nestedness of countries with respect to production stages. Vice versa when countries acted as sellers. The structural organization of countries and productions stages was therefore affected by the nature of the economic transaction.

It is worth noting that even matrices with random entries and optimally reordered rows and columns can exhibit some degree of nestedness~\cite{dominguez2015ranking}. This is especially the case of matrices populated by a large number of ones, since density amplifies the probability of overlapping rows/columns. It is therefore essential to assess whether the values of nestedness computed with the real data statistically significantly differ from the ones that could be obtained using matrices with random entries. To this end, an appropriate null model for the multi-layer network is needed, on which nestedness can be calculated.

A full randomization of the matrix would not represent an appropriate null model, because it would completely destroy the countries' degree distribution in the multi-layer network as well as the degrees of countries in each layer. Besides, the number of connections between pairs of layers would not remain unchanged. On the one hand, an appropriate null model for country-based nestedness would preserve the degree of each country in each layer and across the whole network. On the other, an appropriate null model for transaction-based nestedness would preserve the number of connections linking each pair of layers, i.e., the number of connections linking each layer to itself and to other layers.

Here, we propose two null models that satisfy the above requirements. Fig.~S1 shows a simple example of how the two null models were constructed for testing sellers' country- and transaction-based nestedness. Following~[\citenum{saavedra2011strong}], the first null model (\textit{Model I}) aims to provide a benchmark for assessing the statistical significance of country-based nestedness. For each country $c_i$ and each layer $\alpha$, \textit{Model I} keeps unchanged the number of connections pointing to (for buyers) or departing from (for sellers) country $c_i$ in layer $\alpha$, but randomly reshuffles the layers from (at) which these connections originate (terminate). That is, a given country $c_i$ in layer $\alpha$ will remain involved in the same number of transactions pointing to (departing from) $\alpha$, but these transactions will originate from (terminate at) randomly chosen layers among the $57$ ones (i.e., layer $\alpha$ itself and the remaining others). In terms of the participation matrix, this is equivalent to reshuffling the ones in each row by blocks of columns that share the same importing (exporting) layer. By replicating this procedure for every row, the resulting matrix preserves the global degree of each node as well as the degree of each node in each layer (i.e., the number of links arriving at (departing from) each node in each layer). In summary, \textit{Model I} reshuffles both inter-layer and intra-layer connections while preserving the countries' degree distribution in each layer and across the whole multi-layer network.

The second null model (\textit{Model II}) aims to provide a benchmark for assessing the statistical significance of transaction-based nestedness. To this end, \textit{Model II} randomly assigns countries to production stages, while maintaining the same (number of) connections between pairs of layers (including connections of each layer with itself) and the same countries' degree distributions within each layer as in the real multi-layer network (i.e., it preserves the in-degree distributions for buyers and the out-degree distributions for sellers in each layer, but not across the whole network). To construct such model, for each layer $\alpha$, we kept the connections linking $\alpha$ to itself and to all other layers, but reshuffled the countries at (from) which these connections terminated (originated). That is, the connections to (from) layer $\alpha$ were randomly assigned to countries. In terms of the participation matrix, this is equivalent to swapping rows by blocks of columns defined by the same layer at (from) which connections terminate (depart). For instance, given layer $\alpha$, a block of columns, $\boldsymbol B_{\alpha}$ or $\boldsymbol S_{\alpha}$ for buyers or sellers, respectively, is defined by all pairs of layers exporting to (buyers) or importing from (sellers) layer $\alpha$ (i.e., \{$\alpha\alpha,\alpha\beta, \dots, \alpha\omega$\}). Given any two rows associated with countries $c_i$ and $c_j$, random assignment of countries to transactions would then be obtained by swapping the entire entries of row $c_i$ with the entries of row $c_j$ within block $\boldsymbol B_{\alpha}$ or $\boldsymbol S_{\alpha}$ (i.e., by reassigning to another country the whole set of participations of a given country in transactions in a given block of columns).

In summary, \textit{Model II} randomly reassigns countries to blocks of transactions while preserving the countries' in- or out-degree distribution in each layer (i.e., the distribution of connections arriving at, or departing from, countries in each layer) and the layers' degree distribution (i.e., the distribution of connections between pairs of layers). 

% Results of the null model.
Drawing on the above null models, we generated an ensemble of matrices, both for buyers and for sellers, to evaluate whether nestedness measured using the real data is statistically significantly different from the values one would expect by chance (see methods). Fig.~S2 shows the evolution of country-, transaction-based and total nestedness compared with the values obtained using the appropriate null models. In all cases, nestedness computed in the null models is smaller than nestedness found in the real data (at the $5\%$ significance level). Notice that even in the case of sellers, country-based nestedness appears very close to, and yet remains statistically significantly different from, the values one would expect by chance on a comparable multi-layer network with the same countries' degree distributions within each layer and globally as in the real network.

To further explore the evolution of nestedness, we computed the growth rate of nestedness defined as 

\begin{equation}
G(t+\Delta t) = \frac{N(t+\Delta t)-N(t)}{N(t)},
\end{equation}
where $N(t)$ is nestedness in a given year $t$ and $\Delta t=1$. Fig.~S3A shows the evolution over time of the growth rate of nestedness for buyers. We can observe an oscillatory behavior with a $2$-year period, as well as a remarkable increase in 2005 and a decline in 2008. On the other hand, the growth rate of nestedness for sellers (Fig.~S3B) displays a less clear oscillatory trend, with only small deviations from zero (Fig.~S3B). 

Next, we investigated whether variations in nestedness are associated with economic downturns and, more generally, with the global economic performance of countries. There is a growing body of evidence suggesting that countries' involvement in global value chains is associated with their economic growth and productivity \cite{OECD, worldbank}. Here we relied on our measures of nestedness to capture countries' participation in the international production network. We used data from the World Bank~\cite{WorldBank2017}, and computed the sum of the Gross Domestic Product (GDP) in USD at constant price (2010) of all countries included in our data set. We then examined the relationship between aggregated GDP and the nestedness of the worldwide trade multi-layer network over time. Fig.~S3C shows the buyers' total nestedness as well as country-based and transaction-based nestedness as a function of the GDP. Fig.~S3D, on the other hand, shows the sellers' total, country-based and transaction-based nestedness as a function of GDP. To quantify the association between nestedness $N(t)$ and total GDP, we evaluated the Pearson correlation coefficient, $\rho$, and its $95\%$ confidence intervals (CI) using bootstrapping~\cite{efron1994introduction} (see Table~\ref{t:regression}). Findings suggest that, except for buyers' transaction-based and total nestedness, there is a statistically significant and negative relationship between nestedness and GDP. 

More formally, a simple linear regression model of the relationship between nestedness and GDP in year $t$, $N(t)$ and $GDP(t)$ respectively, can be written as
\begin{eqnarray}\label{eq:nestgdp}
N(t) = \gamma+\beta\,\log_{10}\text{GDP}(t)+\epsilon(t),
\end{eqnarray}
where $\epsilon(t)$ is the residual error term for year $t$ (assumed to be independent of the residuals for other years), and the parameters $\gamma$ (the intercept) and $\beta$ (the regression coefficient) can be estimated through ordinary least-squares (OLS) estimation. The estimated parameters and corresponding standard errors, $p-$values, coefficients of determination $R^2$, and Pearson correlation coefficients $\rho$ for sellers' and buyers' country-based nestedness ($N_c$), transaction-based nestedness ($N_t$) and total nestedness ($N_{total}$) are shown in Table~\ref{t:regression}. The estimated curves are displayed in Fig.~S3.

\begin{table}[h]
\centering
\caption{OLS estimates from regression models and Pearson correlation coefficients.}
\label{t:regression}
\begin{tabular}{lrrrrrr}

Model & $\gamma$ & $\beta$ &  Std. error & $p-$value & $R^2$ & $\rho$ [$95\%$ CI]  \\
\hline
 Buyers' $N_{\text{total}}$ vs. GDP & 47 & 1.16 & 0.50 & 0.03 & 0.53 & 0.29 [0.05,0.79] \\
 Buyers' $N_{\text{c}}$ vs. GDP & 319 & -18.79 & 7.27 & 0.02 & 0.33 & -0.60 [-0.83,-0.24] \\
 Buyers' $N_{\text{t}}$ vs. GDP & 50 & 1.25 & 0.54 & 0.03 & 0.29 & 0.37 [0.08, 0.79]\\
 Sellers' $N_{\text{total}}$ vs. GDP & 173 & -7.71 & 1.60& 0.0003 & 0.64 & -0.71 [-0.91,-0.58] \\
 Sellers' $N_{\text{c}}$ vs. GDP & 251 & -11.28 & 3.20& 0.004 & 0.48 & -0.75 [-0.85,-0.53]  \\
 
 Sellers' $N_{\text{t}}$ vs. GDP & 186 & -8.27 & 1.72 & 0.0003 & 0.63 & -0.86 [-0.92, -0.58]  \\
\hline 
\end{tabular}
\end{table}

% Notice that total nestedness now has a positive correlation with GDP
Overall, these findings suggest that, as GDP increases, nestedness is expected to decline, and vice versa. It may be speculated that the observed negative association between nestedness and GDP is a reflection of the disorder induced in the system by the countries' freedom and ability to tap more economic opportunities and achieve a better allocation of resources. In this sense, higher economic prosperity can be achieved at the expense of an ordered organization of the production system. By contrast, results seem to suggest that a decline in economic prosperity might be associated with more restraints on transactions and stronger constraints on countries' involvement in production stages, thus yielding improvements in global nestedness. While these are simply broad-brush conjectures on associations between economic and structural variables, any attempt to explain any (causal) relationship between nestedness and GDP would clearly require further scrutiny and empirical investigation.

\subsection*{Contribution of countries and industries to nestedness}

To investigate the contribution of individual countries and industries to nestedness, here we propose to evaluate the nestedness that would result subsequent to the reshuffling of the connections involving each country and each industry individually (see methods section). The general idea is that each country and each industry can be associated with an induced variation in country- and transaction-based nestedness respectively, which in turn can be regarded as reflecting the salience of the country or industry to nestedness~\cite{saavedra2011strong}. Indeed this approach would enable us to assess the effects of potential external shocks, such as an unexpected variation in a buyer's or seller's involvement in production stages (here simulated through the reshuffling of the buyer's or seller's connections at the global level), or the unexpected variation in supply or demand of a specific product (here simulated through the reshuffling of connections originating from or terminating at the corresponding layer). In this sense, the effects caused by such reshuffling on global nestedness would shed light on the influence of countries and products/industries on the global structural organization of the worldwide production multi-layer network.

Formally, we define the contributions of country $c_i$ and layer $\alpha$ to country- and transaction-based nestedness as the $Z-$scores calculated, respectively, over an ensemble of multi-layer networks in which the connections involving country $c_i$ and layer $\alpha$ are randomly reshuffled. Formally, we have
\begin{equation}
Z_{c_i} =(N_c-\langle N_{c_i} \rangle)/\sigma
\end{equation}
and
\begin{equation}
Z_{\alpha} = (N_t-\langle N_{\alpha} \rangle)/\sigma,
\end{equation}
\noindent where $N_c$ and $N_t$ are the values of country- and transaction-based nestedness, respectively, calculated on the original data, $\langle N_{c_i} \rangle$ and $\langle N_{\alpha} \rangle$ are the average values of country- and transaction-based nestedness calculated over the ensembles of the matrices resulting from the reshuffling of country $c_i$'s and layer $\alpha$'s connections respectively, and $\sigma$ is the standard deviation of nestedness across these ensembles. Thus, a positive (negative) value of $Z_{c_i}$ would imply a decline (increase) in country-based nestedness resulting from the reshuffling of the connections of country $c_i$ or, alternatively, a positive (negative) contribution of $c_i$ to country-based nestedness. Similarly, a positive (negative) value of $Z_{\alpha}$ implies a negative (positive) effect on transaction-based nestedness resulting from the reshuffling of connections of layer $\alpha$ or, alternatively, a positive (negative) contribution of layer $\alpha$ to transaction-based nestedness.

\begin{figure}[!ht]
\centering
\includegraphics[width=0.98\textwidth]{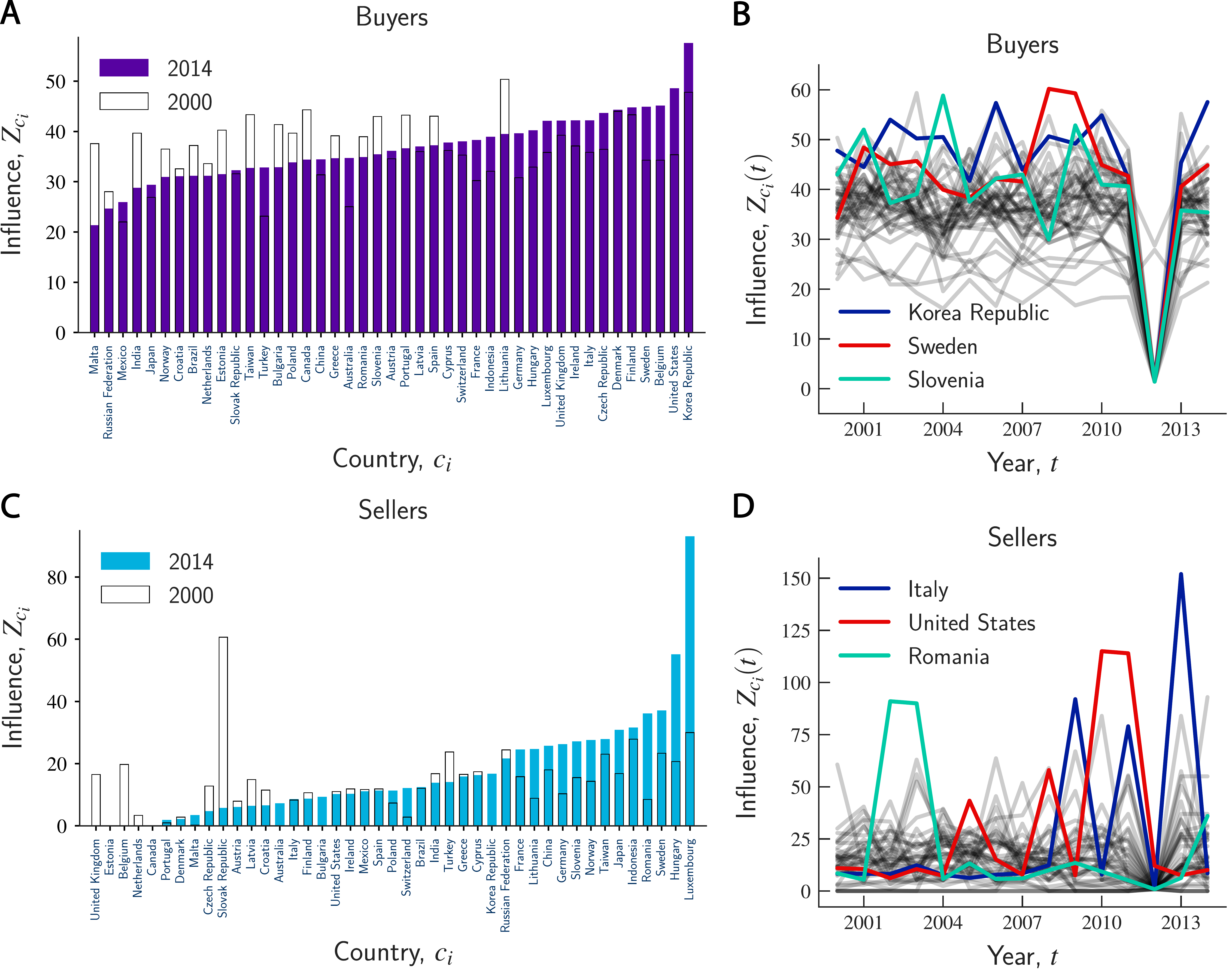}
\caption{Contribution of buyers and sellers to country-based nestedness. A) The bars represent the values of $Z_{c_i}$ for each buyer in 2000 (hollow bar) and 2014 (purple bar). The countries are ranked by the corresponding values of $Z_{c_i}$ in 2014 in an increasing order. All buyers contribute positively to country-based nestedness in both years (i.e., $Z_{c_i}(t)>0$). B) Evolution of $Z_{c_i}(t)$ over the years for buyers. The countries highlighted are the ones associated with the largest variability in contribution to country-based nestedness, i.e., with the largest standard deviation of influence, $\sigma[Z_{c_i}(t)]$, during the observation period. C) Similarly, the bars represent the values of $Z_{c_i}$ for each seller in 2000 (hollow bar) and 2014 (blue bar). The countries are ranked by the corresponding values of $Z_{c_i}$ in 2014 in an increasing order. All sellers have a positive effect on country-based nestedness (i.e., $Z_{c_i}(t)>0$). D) Evolution of $Z_{c_i}(t)$ over the years for sellers. The countries highlighted are the sellers associated with the largest variability in contribution to country-based nestedness, i.e., with the largest standard deviation of influence, $\sigma[Z_{c_i}(t)]$, during the observation period. }
\label{fig:country}
\end{figure}

First, we assessed the contributions of buyers and sellers to country-based nestedness over all years in our data set. Fig.~\ref{fig:country}A shows the influence of individual buyers on country-based nestedness in 2000 and 2014. Findings suggest that an external shock affecting buyers is likely to cause a negative impact on nestedness in both years, i.e., $Z_{c_i} (t) >0$. In addition, we computed the value of $Z_{c_i}(t)$ for each buyer $c_i$ in each year $t$. Fig.~\ref{fig:country}B shows the evolution of $Z_{c_i}(t)$ over time. Fig.~\ref{fig:country}A suggests that Korea Republic, United States, and Belgium are among the importing countries that most contributed to country-based nestedness. Over the whole observation period, these countries are also among those associated with the largest variability in contribution to country-based nestedness (i.e., with the largest standard deviation $\sigma[Z_{c_i}(t)]$). For instance, while Korea Republic is always ranked as one of the suppliers with the largest positive influence on nestedness (Fig.~\ref{fig:country}A), it is also the country whose contribution is characterized by the largest variability over time. 

In the case of sellers, Fig.~\ref{fig:country}C suggests that Luxembourg, Hungary, and Sweden are the exporters characterized by the largest contributions to country-based nestedness in 2014. However, the same ranking was not observed in 2000, unlike the case of buyers that instead almost preserved their ranking in both years. Moreover, it is worth noting that the most influential sellers, ranked as top contributors to country-based nestedness, are not among the exporters that experienced the largest variability in such contribution over the years. Fig.~\ref{fig:country}D shows that Italy, United States, and Romania are the exporting countries associated with the largest standard deviation of $Z_{c_i}(t)$ (i.e., $\sigma[Z_{c_i}(t)]$), thus suggesting different production trajectories for buyers and sellers.

\begin{figure}[!ht]
\centering
\includegraphics[width=0.98\textwidth]{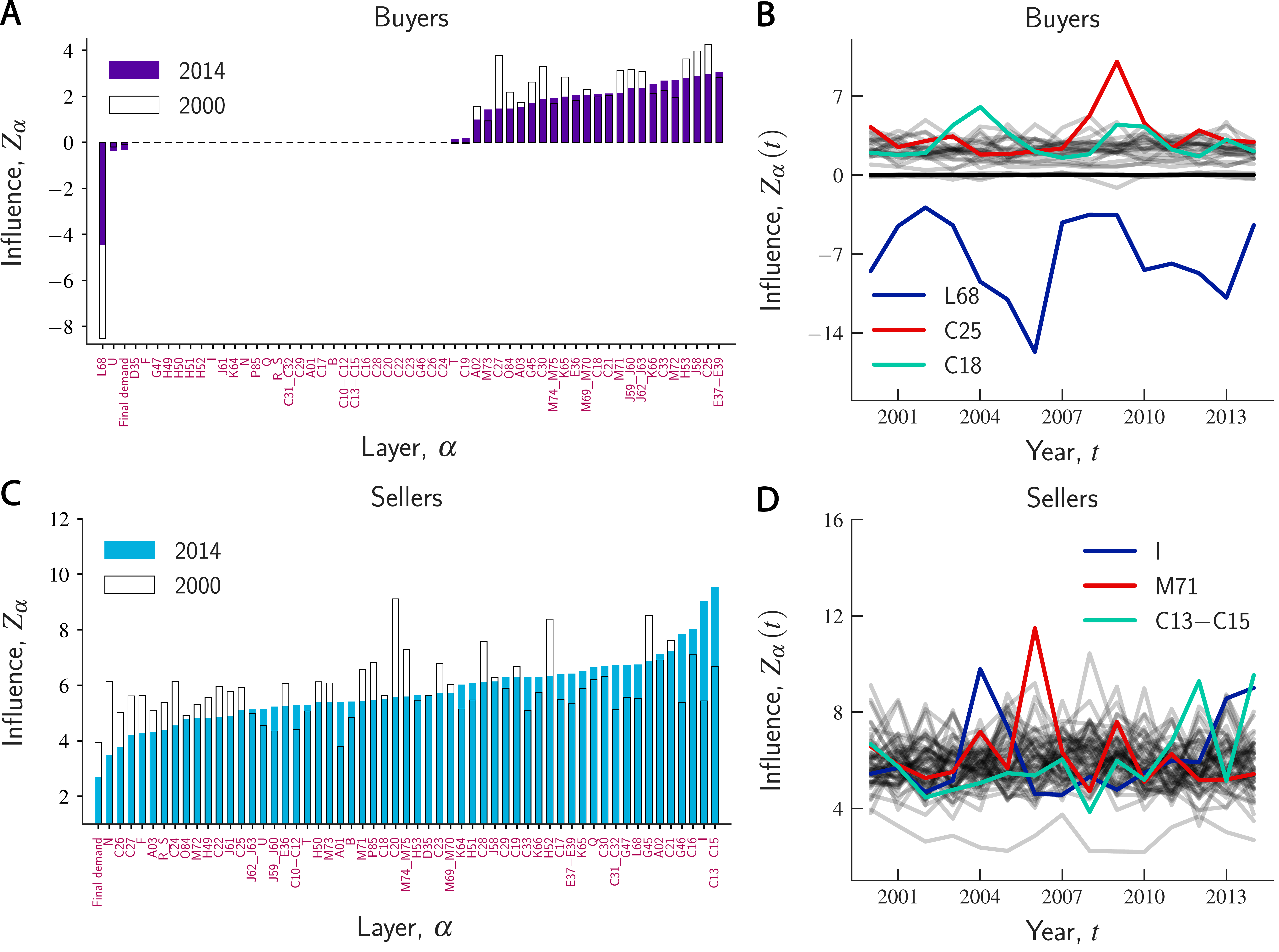}
\caption{Contribution of industries to buyers' and sellers' transaction-based nestedness. A) The bars represent the values of $Z_{\alpha}$ for each layer $\alpha$ in 2000 (hollow) and 2014 (purple). The layers are ranked by the corresponding values of $Z_{\alpha}$ in 2014 in an increasing order. Almost half of the layers have a positive influence on buyers' transaction-based nestedness, (i.e., $ Z_{\alpha}(t)>0$), while several layers have no influence and a minority are characterized by a negative influence. B) Evolution of $Z_{\alpha}(t)$ over the years for buyers. The layers highlighted are the industries associated with the largest variability in contribution to buyers' transaction-based nestedness, i.e., with the largest standard deviation of influence, $\sigma[Z_{\alpha}(t)]$, during the observation period. C) All layers have a positive effect on the sellers' transaction-based nestedness (i.e., $Z_{\alpha}(t)>0$) both in 2000 (hollow bar) and 2014 (blue bar). D) Evolution of $Z_{\alpha}(t)$ over the years for sellers. The layers highlighted are the industries associated with the largest variability in contribution to sellers' transaction-based nestedness, i.e., with the largest standard deviation of influence, $\sigma[Z_{\alpha}(t)]$, during the observation period.}
\label{fig:layer}
\end{figure}

To evaluate the contribution of individual economic activities to transaction-based nestedness in the worldwide production network, we followed a similar procedure to the one used for investigating the contributions of countries to country-based nestedness. In this case, we calculated the values of $Z_{\alpha}$ by randomly allocating to countries the connections departing from (sellers) or arriving at (buyers) each layer at a time (see methods). Fig.~\ref{fig:layer}A shows the effect of individual layers on the buyers' transaction-based nestedness in 2000 and 2014. Results suggest that the economic industries that have the largest influence on the buyers' transaction-based nestedness are the ones related to: sewerage; waste collection, treatment and disposal activities; materials recovery; remediation activities and other waste management services (E37-E39); manufacture of fabricated metal products, except machinery and equipment (C25); and publishing activities (J58). Notice that real estate activities (L68) have the largest negative influence on nestedness in both years.

Just as with the contributions of countries, we measured $Z_{\alpha}(t)$ for each layer $\alpha$ in each year $t$, and uncovered the layers with the largest variability in contribution to buyers' transaction-based nestedness over time, i.e., the layers with the greatest standard deviation $\sigma[Z_{\alpha}(t)]$. Fig.\ref{fig:layer}B shows that the production sectors with the largest positive and negative contributions to nestedness are also among the ones with the greatest variability in contribution to nestedness over the years.

Finally, we investigated the contributions of individual layers to sellers' transaction-based nestedness. Fig.~\ref{fig:layer}C shows the extent to which the reshuffling of connections involving each industry at a time affects the sellers' transaction-based nestedness. In this case, all layers have a positive effect on transaction-based nestedness (i.e., $Z_{\alpha}(t)>0$). Fig.~\ref{fig:layer}D also suggests that the production sectors associated with the largest contributions to sellers' nestedness are among the ones associated with the greatest variability in such contribution over the years (e.g., manufacture of textiles, wearing apparel and leather products (C13-C15) and accommodation and food service activities (I)).

\section*{Discussion}
In this work, we argued that the bipartite network of connections between countries and industries cannot fully capture the economic interactions among countries underpinning the worldwide production network and the global value chain. In this study, we proposed to formalize the interactions between countries and between industries by constructing a multi-layer exchange network in which each layer represents an industry and the nodes represent the trading countries. In this network, countries can be connected within and across layers, and the same country can be involved in multiple stages of the global value chain as well as in transactions within the same industry. Thus, the network includes intra- and inter-layer connections that may depart from and point to any country (i.e., inter-layer connections from and to the same country and self-loops of length one are allowed). Based on the multi-layer network, we built two participation matrices, one for buyers and the other for sellers, in which rows are the (buying or selling) countries and columns are all possible ordered combinations of any two layers. We then used these matrices to compute both country-based and transaction-based nestedness, each from the perspective of buyers and sellers.

We showed that, as typically occurs within ecological networks, the structural organization of the participation matrices of buyers and sellers displays a nested signature. Such nested structure is significantly statistically different from the structure one would expect in a multi-layer null model in which connections are randomly reshuffled while the countries' or layers' degree distributions are kept unaltered. Our results suggested that, while sellers' nestedness remained constant over the years, buyers' country-based nestedness fluctuated between 2003 and 2011. We also uncovered associations between nestedness and GDP, and found that, except for buyers' transaction-based nestedness, an increase in GDP is associated with a decline in nestedness. 

Moreover, we investigated the contributions of individual countries and individual economic industries to country- and transaction-based nestedness, respectively, by reshuffling the connections of one country or one layer at a time, and by comparing the resulting nestedness with the original one. To this end, we computed $Z-$scores over an ensemble of multi-layer networks, both for buyers and sellers. Results indicated that, while all countries exerted a positive contribution to buyers' country-based nestedness, a minority of countries played no significant role in sustaining nestedness. These are the countries whose reshuffled connections did not induce any change in nestedness and therefore whose $Z-$scores did not differ from zero. We further explored variations over time of countries' contributions to country-based nestedness, and found that the top-ranked contributors tend to be precisely the suppliers associated with the largest variation of such contributions over the years. It is worth noting that, among the top contributors (both buyers and sellers), there are also relatively smaller, developing and emerging countries (e.g., Luxembourg, Korea Republic, and the Czech Republic among the buyers, and  Luxembourg and Hungary among the sellers) as well as larger and more advanced countries (e.g., the United States and Belgium among the buyers and Japan and Germany among the sellers). Similarly, among the countries that contributed least to nestedness are not only smaller East European countries (e.g., Croatia among the buyers and Estonia and Czech Republic among the sellers) and developing countries (e.g., India among the buyers), but also large advanced countries (e.g., Japan among the buyers and the United Kingdom and Belgium among the sellers). These findings might therefore suggest that a country's contribution to nestedness is not necessarily correlated with its economic size. 
Results also indicated that, while all industries had a positive effect on the sellers' transaction-based nestedness, a number of them (e.g., mining and quarrying (B) and human health and social work activities (Q)) played no role in sustaining the buyers' transaction-based nested organization of the international production multi-layer network.

We believe that our multi-layer network perspective can shed a new light on the structural foundations and dynamics of competition between countries and industries in the global value chain, on the global system's vulnerability to exogenous shocks, on the contribution that each component of the system can make to global nestedness, and ultimately on the role that nestedness can play in enhancing or inhibiting economic growth over time. Because a variety of empirical domains, from biological to technological and social ones, can be characterized as complex networks in which relationships have a multi-layer representation, our approach also has important implications beyond international trade, and can help gain a better understanding of the structural organization, stability, and growth mechanisms of a number of different systems.

\section*{Methods}
\subsection*{The data set}
We used data from the WIOD, which is connected to a project funded by the European Commission as part of the $7^{th}$ Framework Programme with the aim of developing new databases, accounting frameworks and models to increase our understanding of the dynamic interrelatedness of countries and industries. 
The core of the database is a set of harmonized supply and use tables, as well as data on international trade in goods and services. These two sets of data have been integrated into sets of inter-country (world) input-output tables. For further information, please refer to Data Description in the ``Supplementary Information'' section.

\subsection*{Reordering of columns and rows}
To calculate nestedness, the rows (countries) of the participation matrix were sorted in decreasing order based on the number of transactions (i.e., pairs of layers) in which countries were involved. Similarly, the columns of the participation matrix were sorted in decreasing order by number of countries that shared involvement in transactions. 

\subsection*{Null models}
To test the hypothesis that countries and stages of production have a nested organization in the multi-layer network, we assessed  whether the value of nestedness measured using our data is statistically significantly different from the value one would obtain by random expectation. To this end, we drew on, and extended, the ideas proposed by Saavedra {\it et al.}~\cite{saavedra2011strong}, and constructed appropriate null models (I and II) for the multi-layer network. We also used these models to assess the influence of individual countries and individual industries upon country- and transaction-based nestedness, respectively. All the results were obtained based on an ensemble of $1,000$ realizations of the null models. We assumed as statistically significant (at the 5\% level) all observed values of country-based, transaction-based, and total nestedness that lie outside the 95\% confidence interval of nestedness obtained using the corresponding null models over $1,000$ replicates.

\textbf{Null model I.} To test the statistical significance of country-based nestedness, we followed the approach proposed in [\citenum{saavedra2011strong}] and constructed \textit{Model I}. In particular, we randomized the rows of the participation matrix by blocks of columns defined by the common importing (for buyers) or exporting (for sellers) layer, while keeping unchanged the number of ones in each row (i.e., the countries' degree distribution across the whole multi-layer network) and also preserving the degree of each node in each layer (i.e., the intra-layer countries' degree distribution). For example, if node $c_i$ has two links pointing to (departing from) layer $\alpha$ and one departing from (pointing to) layer $\beta$ and the other from (to) layer $\gamma$, in \textit{Model I} node $c_i$ would still have two links pointing to (departing from) layer $\alpha$ but the origin (destination) of these links would be randomly reassigned to any other layer among all available ones (i.e., including layer $\alpha$ itself). Thus, for each country (buyer or seller) \textit{Model I} preserves the total number of links ending at (buyers) or departing from (sellers) layer $\alpha$. As a result of this constraint, the total number of links of each node across the whole network is also preserved. We used the same model to investigate the influence of individual countries on country-based nestedness. That is, for each country $c_i$ we randomized (subject to the above constraints) the corresponding row in the participation matrix, and then computed the $Z-$score over the ensemble produced by the realizations of the model.

\textbf{Null model II.} To test the statistical significance of transaction-based nestedness, we followed an approach similar to the one outlined above and constructed \textit{Model II}. In particular, we randomized the rows of the participation matrix by blocks of columns defined by the common importing (for buyers) or exporting (for sellers) layer, while preserving not only the countries' (in- or out-) degree distribution in each layer of the multi-layer network (i.e., the intra-layer in- or out-degree distribution), but also the layers' degree distribution (i.e., the number of connections from one layer to another and/or to itself did not change). This is equivalent to swapping the connections of country $c_i$ pointing to (originating from) layer $\alpha$ with the connections of another country $c_j$ pointing to (originating from) the same layer $\alpha$. We also used \textit{Model II} to investigate the impact of individual industries to transaction-based nestedness. To this end, for each individual layer $\alpha$, we randomized (subject to the above constraints) the corresponding block of columns (i.e., all columns in which layer $\alpha$ is the destination or origin of connections) in the participation matrix, and then computed the $Z-$score over the ensemble produced by the realizations of the model.

\section*{Acknowledgements}
L.G.A.A. acknowledges FAPESP (Grant No. 2016/16987-7) for financial support.
F.A.R. acknowledges CNPq (Grant No. 305940/2010-4) and FAPESP (Grants No. 2016/25682-5 and grants 2013/07375-0). Y. M. acknowledges support from the Government of Arag\'on, Spain through a grant to the group FENOL, by MINECO and FEDER funds (grant FIS2014-55867-P) and by the European Commission FET-Proactive Project Dolfins (grant 640772). 

\section*{Author contributions statement}
%All authors contributed equally to this work.
L. G. A. A., G. M., I. C., F. A. R., P. P. and Y. M. designed research, L. G. A. A., G. M., I. C., F. A. R., P. P. and Y. M. performed research, L. G. A. A., G. M., I. C., F. A. R., P. P. and Y. M. analyzed the data and results, and L. G. A. A., G. M., I. C., F. A. R., P. P. and Y. M. wrote the manuscript. All authors reviewed and approved the manuscript.
\section*{Competing interests} The authors declare no competing interests. 

\section*{Additional information}

\setcounter{figure}{0}
\makeatletter 
\renewcommand{\thefigure}{S\@arabic\c@figure}
\renewcommand{\thetable}{S\@arabic\c@table}
\newpage
\section*{Data description}

Table~\ref{t:activities} the $56$ NACE Rev.2 economic activities divisions included in the WIOD, and Table~\ref{t:countries} shows the list of $43$ countries (excluding the Rest of the World).

\begin{longtable}{|p{3cm}|p{8cm}|}

\hline \multicolumn{1}{|c|}{\textbf{NACE Rev. 2 Division}} & \multicolumn{1}{c|}{\textbf{Economic activity description}} \\ \hline 
\endfirsthead

\multicolumn{2}{c}%
{{\bfseries \tablename\ \thetable{} -- continued from previous page}} \\
\hline \multicolumn{1}{|c|}{\textbf{NACE Rev. 2 Division}} & \multicolumn{1}{c|}{\textbf{Economic activity description}} \\ \hline 
\endhead

\hline \multicolumn{2}{|r|}{{Continued on next page}} \\ \hline
\endfoot

\hline \hline
\endlastfoot

A01 & Crop and animal production, hunting and related service activities\\ \hline
A02    &    Forestry and logging\\ \hline
A03    &    Fishing and aquaculture\\ \hline
B    &    Mining and quarrying\\ \hline
C10-C12    &    Manufacture of food products, beverages and tobacco products\\ \hline
C13-C15    &    Manufacture of textiles, wearing apparel and leather products\\ \hline
C16    &    Manufacture of wood and of products of wood and cork, except furniture; manufacture of articles of straw and plaiting materials\\ \hline
C17    &    Manufacture of paper and paper products\\ \hline
C18    &    Printing and reproduction of recorded media\\
C19    &    Manufacture of coke and refined petroleum products \\ \hline
C20    &    Manufacture of chemicals and chemical products \\ \hline
C21    &    Manufacture of basic pharmaceutical products and pharmaceutical preparations\\ \hline
C22    &    Manufacture of rubber and plastic products\\ \hline
C23    &    Manufacture of other non-metallic mineral products\\ \hline
C24    &    Manufacture of basic metals\\ \hline
C25    &    Manufacture of fabricated metal products, except machinery and equipment\\ \hline
C26    &    Manufacture of computer, electronic and optical products\\ \hline
C27    &    Manufacture of electrical equipment\\ \hline
C28    &    Manufacture of machinery and equipment n.e.c.\\ \hline
C29    &    Manufacture of motor vehicles, trailers and semi-trailers\\ \hline
C30    &    Manufacture of other transport equipment\\ \hline
C31\_C32    &    Manufacture of furniture; other manufacturing\\ \hline
C33    &    Repair and installation of machinery and equipment\\ \hline
D35    &    Electricity, gas, steam and air conditioning supply\\ \hline
E36    &    Water collection, treatment and supply\\ \hline
E37-E39    &    Sewerage; waste collection, treatment and disposal activities; materials recovery; remediation activities and other waste management services \\ \hline
F        & Construction\\ \hline
G45    &    Wholesale and retail trade and repair of motor vehicles and motorcycles\\ \hline
G46    &    Wholesale trade, except of motor vehicles and motorcycles\\ \hline
G47    &    Retail trade, except of motor vehicles and motorcycles\\ \hline
H49    &    Land transport and transport via pipelines\\ \hline
H50    &    Water transport\\ \hline
H51    &    Air transport\\ \hline
H52    &    Warehousing and support activities for transportation\\ \hline
H53    &    Postal and courier activities\\ \hline
I    &    Accommodation and food service activities\\ \hline
J58    &    Publishing activities\\ \hline
J59\_J60    &    Motion picture, video and television programme production, sound recording and music publishing activities; programming and broadcasting activities\\ \hline
J61    &    Telecommunications\\ \hline
J62\_J63    &    Computer programming, consultancy and related activities; information service activities\\ \hline
K64    &    Financial service activities, except insurance and pension funding\\ \hline
K65    &    Insurance, reinsurance and pension funding, except compulsory social security\\ \hline
K66    &    Activities auxiliary to financial services and insurance activities\\ \hline
L68    &    Real estate activities\\ \hline
M69\_M70    &    Legal and accounting activities; activities of head offices; management consultancy activities\\ \hline
M71    &    Architectural and engineering activities; technical testing and analysis\\ \hline
M72    &    Scientific research and development\\ \hline
M73    &    Advertising and market research\\ \hline
M74\_M75    &    Other professional, scientific and technical activities; veterinary activities\\ \hline
N    &    Administrative and support service activities\\ \hline
O84    &    Public administration and defense; compulsory social security\\ \hline
P85    &    Education\\ \hline
Q    &    Human health and social work activities\\ \hline
R\_S    &    Other service activities\\ \hline
T    &    Activities of households as employers; undifferentiated goods- and services-producing activities of households for own use\\ \hline
U    &    Activities of extraterritorial organizations and bodies\\ \hline

\hline
\caption{Economic activities in WIOD 2016 Release}\label{t:activities}
\end{longtable}

\begin{table}[h]
\center
\begin{tabular}{|p{12cm}|}
\hline
\\
{\bf Country name (ISO Alpha-3 Code)} \\
\\
\hline
\\
Australia (AUS), Austria (AUT), Belgium (BEL), Bulgaria (BGR), Brazil (BRA), Canada (CAN), Switzerland (CHE), China (CHN), Cyprus (CYP), Czech Republic (CZE), Germany (DEU), Denmark (DNK), Spain (ESP), Estonia (EST), Finland (FIN), France (FRA), United Kingdom (GBR), Greece (GRC), Croatia (HRV), Hungary (HUN), Indonesia (IDN), India (IND), Ireland (IRL), Italy (ITA), Japan (JPN), Korea, Rep. (KOR), Lithuania (LTU), Luxembourg (LUX), Latvia (LVA), Mexico (MEX), Malta (MLT), Netherlands (NLD), Norway (NOR), Poland (POL), Portugal (PRT), Romania (ROU), Russian Federation (RUS), Slovak Republic (SVK), Slovenia (SVN), Sweden (SWE), Turkey (TUR), Taiwan (TWN), United States (USA)\\

\hline
\end{tabular}
\caption{Countries in WIOD 2016 Release}\label{t:countries}
\end{table}

\newpage

\section*{Supplementary figures}
\begin{figure}[!h]
\centering
\includegraphics[width=0.9\textwidth]{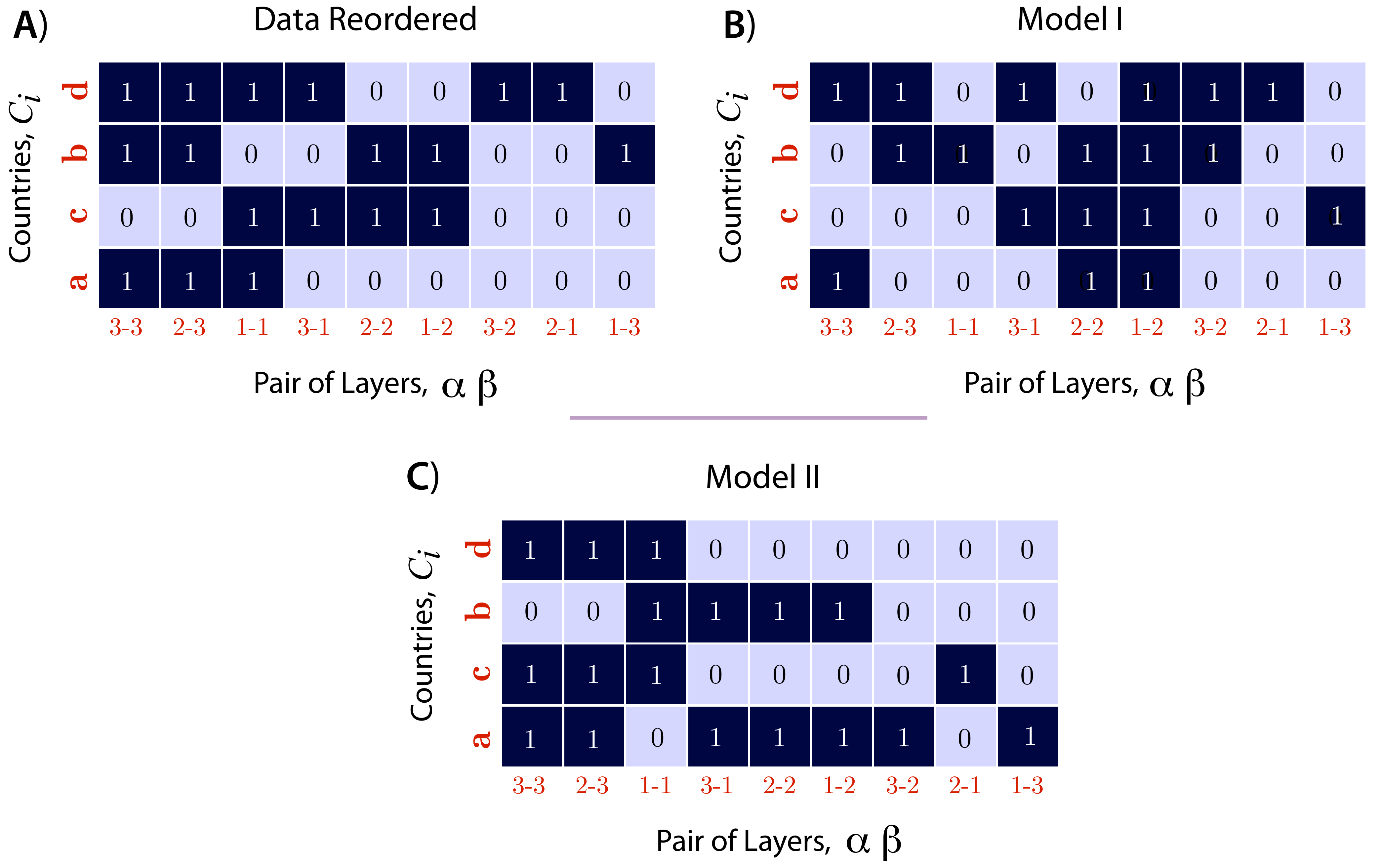}
\caption{{A simple example of the two null models for testing sellers' country-based and transaction-based nestedness. A) The original participation matrix in which the rows are four exporters and the columns are nine different combinations of three layers (industries). The matrix has been reordered by row/column degree (i.e., number of ones). B) \textit{Model I}. For each row at a time, columns are reshuffled within (three) blocks defined by the common exporting layer (i.e., block [1-1, 1-2, 1-3], block [2-1, 2-2-, 2-3], and block [3-1, 3-2, 3-3]). Within each block, the ones are randomly reshuffled, while the total number of ones is preserved. For example, in block [1-1, 1-2, 1-3] the one moves from column [1-1] in the original data to column [1-2] in the null model. Notice that this model randomly reassigns end layers to the starting layers of transactions while keeping the out-degree of each country unchanged both within each layer and across the whole network. C) \textit{Model II}. For each layer at a time, rows are reshuffled by (three) blocks of columns defined by the common exporting layer. Within each block, the entire rows are randomly swapped, while the total number of ones is preserved. For example, in block [1-1, 1-2, 1-3] the entries of row $b$ have been swapped with the entries of row $a$. Notice that this model randomly reassigns countries to transactions starting at a given layer while keeping the degree distribution of layers across the whole network. Moreover, the out-degree distribution of countries is preserved within each layer, but not across the whole network. 
}}
\label{fig:nullmodel}
\end{figure}

\begin{figure}[!h]
\centering
\includegraphics[width=0.95\textwidth]{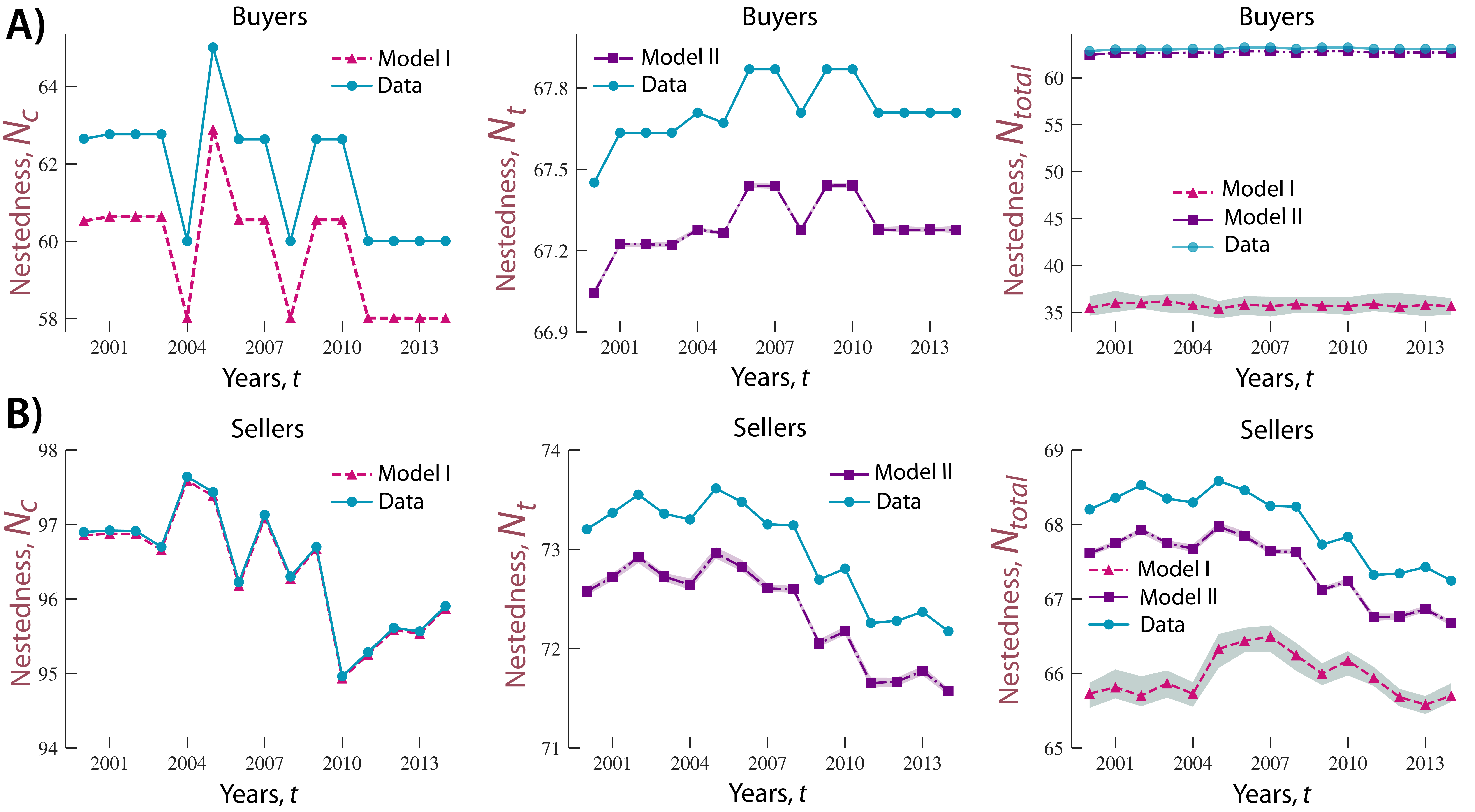}
\caption{{ Null models compared with real data. Buyers' (A) and sellers' (B) nestedness compared with values obtained with the null models. The nestedness of countries is always higher than the values obtained by using \textit{Model I}, that is by using a multi-layer network in which connections are randomly reshuffled but that preserves the same node degree distribution (i.e., the global and intra-layer degree distributions) as in the real network. The nestedness of production stages (transactions) is also higher than the one found using \textit{Model II}, that is by using a multi-layer network in which connections are randomly reshuffled among the countries in the same layer, but that preserves the countries' intra-layer degree distribution (in-degree distribution for buyers and out-degree distribution for sellers) for each layer as well as the layers' degree distribution. All observed values of country-based, transaction-based, and total nestedness are outside the 95\% confidence interval of nestedness obtained using the corresponding null models over $1,000$ replicates. Notice that the thin shaded areas reflecting confidence intervals are explained by the small variation in values obtained using null models based on very dense participation matrices (i.e., percentage of ones in the matrix is $\approx 90\%$).} 
}
\label{fig:nullmodel_evolution}
\end{figure}

\begin{figure}[!h]
\centering
\includegraphics[width=0.9\textwidth]{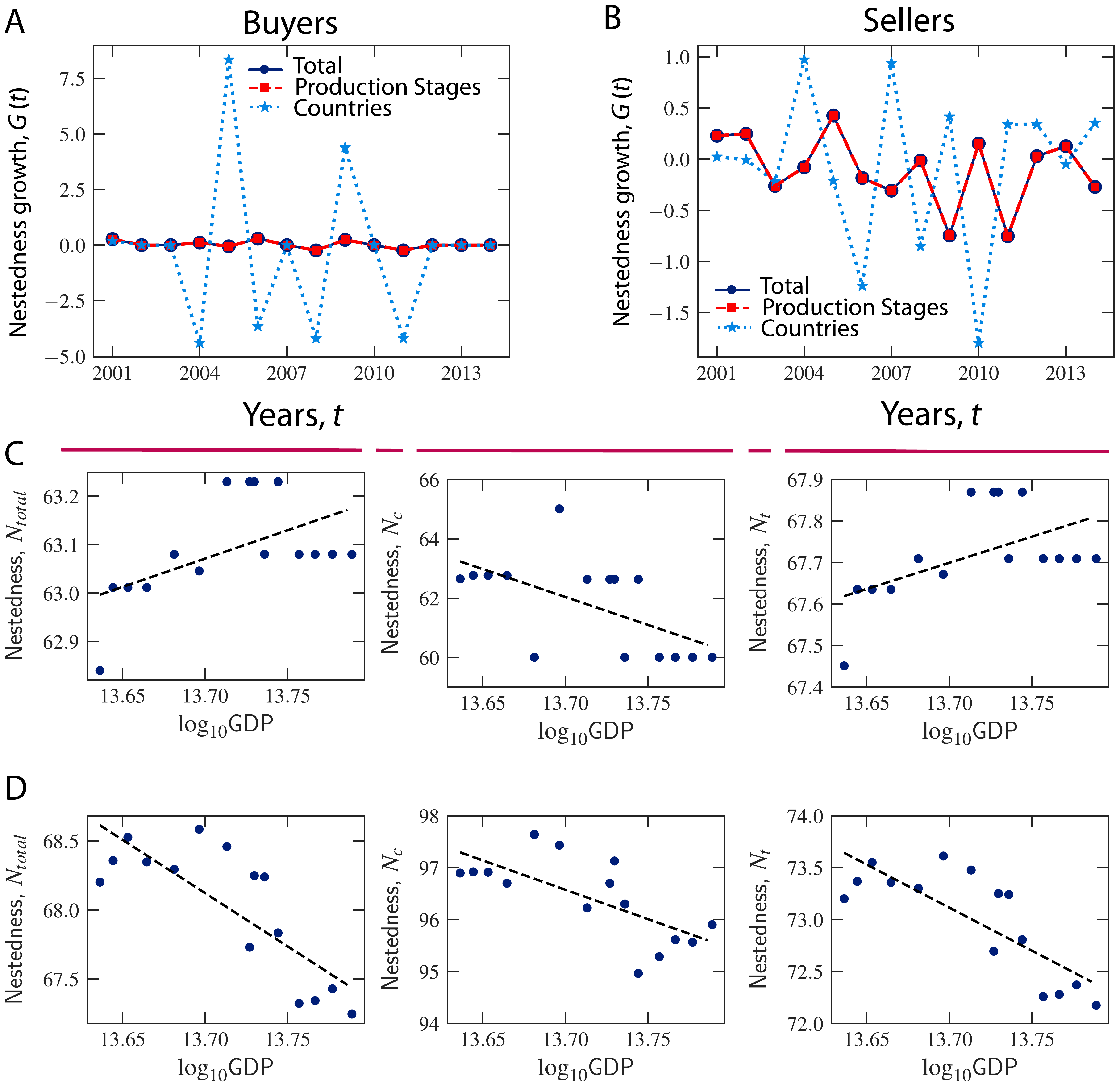}
\caption{Variations in nestedness and relationship between nestedness and GDP. A) Growth in global nestedness calculated over the years from 2000 to 2014 from the buyers' perspective. B) Growth in global nestedness calculated over the years from 2000 to 2014 from the sellers' perspective. C) Relationship between buyers' nestedness and total GDP. D) Relationship between sellers' nestedness and total GDP. The black dashed lines are the OLS fit to Eq.~7 (see main text). See Table~1 in the main text for the estimated regression coefficients.}
\label{fig:nestednessgdp}
\end{figure}

\end{document}